\documentclass{article}
\usepackage{amsmath,amssymb}
\usepackage{tikz}
\usepackage{xfrac}
\usepackage{mathtools, cancel}
\usepackage{dsfont, titlesec}

\numberwithin{equation}{section}
\numberwithin{figure}{section}

\def\a{{\alpha}}
\def\b{{\beta}}
\def\g{{\gamma}}
\def\w{{\omega}}
\def\t{{\tau}}
\def\th{{\theta}}
\def\s{{\sigma}}

\def\thao{{\theta_{A_1}}}
\def\thaa{{\theta_{A_2}}}
\def\thbo{{\theta_{B_1}}}
\def\thbb{{\theta_{B_2}}}

\def\thav{{\theta_{av}}}
\def\thbv{{\theta_{bv}}}
\def\thab{{\theta_{ab}}}
\def\thv{{\theta_{v}}}
\def\tha{{\theta_{a}}}
\def\thb{{\theta_{b}}}
\def\thm{{\psi_{-}}}

\def\G{{\Gamma}}

\def\R{{\mathbb{R}}}

\def\Z{{\mathbb{Z}}}

\def\M{{\mathcal{M}}}
\def\F{{\mathcal{F}}}
\def\Mg{({\mathcal{M}},g)}
\def\Mf{({\mathcal{M}},{\mathcal{F}})}

\def\c{{\mathcal{C}}}

\def\Rc{{\mathcal{R}}}

\def\O{{\mathcal{O}}}
\def\go{\gamma_{\mathcal{O}}}

\def\Ko{K^{\O}}

\def\Oa{{\O_A}}
\def\Ob{{\O_B}}

\def\da{{d^{(A)}}}
\def\db{{d^{(B)}}}
\def\ma{{m^{(A)}}}
\def\mb{{m^{(B)}}}
\def\va{{v^{(A)}}}
\def\vb{{v^{(B)}}}

\def\lp{{J^-(p)}}
\def\le{{J^-(p_E)}}

\def\E{{\mathcal{E}}}

\def\W{{\Omega}}
\def\Wo{{\W^{\O}}}
\def\L{{\Lambda}}
\def\Lo{{\L^{\O}}}

\def\P{{\mathbb{P}}}
\def\Po{{\P^{\O}}}

\def\UP{{\cal UP}}

\def\Pth{{\cal P}}
\def\Ptho{{\cal P}^{\O}}

\def\Pthop{{(\W^{\O}_p,\L^{\O}_p,\P^{\O}_p)}}

\newcommand\beq{\begin{equation}}
\newcommand\eeq{\end{equation}}
\newcommand\bea{\begin{eqnarray}}
\newcommand\eea {\end{eqnarray}}
\newcommand\ba{\begin{array}}
\newcommand\ea {\end{array}}

\newcommand\bd{\begin{description}}
\newcommand\ed {\end{description}}
\newcommand\ben{\begin{enumerate}}
\newcommand\een{\end{enumerate}}

\def\nn{{\nonumber} }

\begin{document}

\begin{center}
   \baselineskip=16pt
   \begin{LARGE}
      \textsl{Observer Dependence in Consistent Histories and General Relativity}
   \end{LARGE}
   \vskip 1cm
      Yousef Ghazi-Tabatabai
   \vskip .2cm
   \begin{small}
      \textit{yousef.tabatabai@cantab.net}
    \end{small}
   \vskip 1cm
\end{center}


\begin{abstract}
\noindent Observer dependence is central to Quantum Mechanics; in particular if we associate consistent sets with observer worldviews we can think of the Consistent Histories (CH) interpretation as a formalisation of observer dependence. However we lack a theory of the observers themselves. In this paper we begin by building such a theory within General Relativity (GR), and find that this framework bears close mathematical resemblance to the treatment of consistent sets in CH. We therefore adapt the CHSH framework to identify the `gap' between the classical and quantum theories. We find that the CHSH argument does not hold on a curved background, so that the inequality may be violated by classical theories.
\end{abstract}

\section{Introduction}

\subsection{Observer Dependence}

Observer dependence is a central feature of Quantum Mechanics (QM), closely related to the Copenhagen `measurement problem'  and the idea of `consistent sets' in the histories formalism. Indeed, we are able to think of many QM features in terms of observer independence.

For example, if we regard Schr\"{o}dinger's cat as an observer (if necessary by replacing it with a physicist) we would not expect it to perceive itself as being in a superposition. The controversial `dead and alive' superposition is then a feature of the second observer's `worldview'\footnote{To use the term employed by Isham when applying topos methods to construct logical structures in Consistent Histories \cite{isham:toposcs}.}. The key consideration is that the cat is `in a box' and unable to communicate with the second observer, in other words they are spacelike. Generalising, two spacelike observers may perceive the same event differently, each observer placing the event within the context of its own knowledge and worldview. Further, the details of the measurement process may differ between the two observers, a `detail' which should not be ignored. Indeed key thought (and actual) experiments are built upon such spacelike observers with differing worldviews, including the EPR argument \cite{epr}, the related Bell and CHSH inequalities \cite{bell:1964, chsh} and the realisation of the later thought experiment by Aspect, Dalibard \& Roger \cite{aspect}. Generalising further, we might associate the `many worldviews' of Consistent Histories (CH) with the worldviews of `many observers' at different points in their respective times. 

However the study of observer dependence in QM rarely involves the any model of the observers themselves, the way in which such observers formulate their respective `worldviews', or how the relation between observers corresponds to the relation between the ways in which they perceive a jointly observed phenomenon. We will explore these questions, taking Consistent Histories as our starting point.

\subsection{Consistent Histories}

Building on the work of Dirac \cite{dirac:histories} and Feynman \cite{feynman:1948}, the \emph{histories approach} was pioneered by Omnes \cite{omnes:1988}, Griffiths \cite{griffiths:1984,griffiths:1996,griffiths:1998}, Gell-Mann and Hartle \cite{gellman:1990, hartle:1992}. Moving away from `state-vectors', this approach focuses on the set $\UP$ of \emph{histories} of a system, whose dynamics is described by the \emph{decoherence functional}. Central to the approach are the \emph{consistent sets}, partitions of $\UP$ upon which the decoherence functional can be reduced to a standard probability measure.  More concretely, for every consistent set $\W$ we are able to construct a standard probability theory $(\W,\L,\P)$ where $\L\subseteq \W$ is the event algebra and $\P$ a probability measure derived from the decoherence functional in the standard manner. 

Consistent sets are the histories analogue of experimentally observable events\footnote{In a given system, every Copenhagen observable event will correspond to a consistent set, however the converse may not be true in general.} and are the focus of the \emph{Consistent Histories} (CH) interpretation \cite{griffiths:1996,griffiths:1998}. CH allows us to simultaneously assign truth values to all propositions within a single consistent set, and to apply Boolean logic within this set in a manner consistent with the truth valuation. However two elements which are not members of the same consistent may not participate in a Boolean logic, and can not simultaneously receive truth values. Algebraically this is reflected by the fact that the consistent sets generate Boolean subalgebras of the orthoalgebra\footnote{Or equivalently the boolean manifold $\UP$, which may be a more intuitive structure for CH} $\UP$.

With reference to the `many worlds' expression, Isham \cite{isham:toposcs} refers to the consistent sets as `many worldviews', and applies topos techniques to bring them together into a single logical structure including global truth valuation. This work has been built on by Isham and D\"{o}ering to provide a more general application of topos techniques to quantum physics \cite{isham:topos1, isham:topos2, isham:topos3, isham:topos4, isham:2010, isham:2012}.

\subsection{Searching for a Formal Theory}

Now every observation (or Copenhagen measurement) by every observer corresponds to a consistent set, and leads to one element of that set being valued as `true' and the others as `false' (analogous to state-vector reduction). However the converse may not be true in general, and we might restrict to the less well defined notion of `observable' consistent sets (or the even less well defined notion of `actually observed' consistent sets) so that the correspondence becomes one-to-one. Assuming we are able to do this successfully, and identifying observers worldviews with their (potential) observations, we can view CH as a formalisation of observer dependence in QM. 

However, as noted above we lack a formal framework describing the observers themselves. Focusing on `spacetime observers', we might look to the spacetime structure to provide a framework for `putting together' the observers and their worldviews. However we lack a successful theory of Quantum Gravity; further the need to interpret such a theory in terms of observers risks circularity. It therefore seems constructive to begin with General Relativity (GR), and a formulation of observer dependence in an entirely classical theory.

\subsection{Goals and Outline of this Paper}

Despite the fact that the term `relativity' itself refers to observer dependence, historically this question has not received the same degree of attention in GR as it has in QM, perhaps because it has never been seen as a `problem'. Likewise, the related concept of measurement has received less scrutiny in GR. 

In this paper we make an initial exploration of observer dependence in GR, examining the observers themselves, the way in which they construct their worldviews and how the relations between these worldviews corresponds to the relations between the observers within spacetime, particularly as touches the measurement of jointly observed phenomena. We will have an eye on the potential generalisation to quantum theory, and to this end we will make frequent  comparisons with CH, aiming to identify the `gap' between the classical and quantum theories where possible. To this end we will devote particular attention to the EPR-Bell-CHSH argument, which was designed to identify and test the gap between classical and quantum behaviour.

In section $2$ we explore and formalise the way in which an observer and its worldviews might be defined in terms of standard GR structures. In $2.1$ we formalise the concept of an observer, following and extending the standard approach used in GR. In $2.2$ we build a formal theory of our observers' worldviews, and in $2.3$ we compare these structures with the worldviews of CH. In section $3$ we adapt the CHSH framework to GR. In $3.1$ we carefully outline the CHSH thought experiment on a curved background, in $3.2$ we determine the general dynamics of the experiment and in $3.3$ we introduce simplifying assumptions which allow us to make a direct comparison with QM. We discuss the results in $3.4$. In section $4$ we take initial steps toward constructing a global logical structure, relating the worldviews directly to spacetime in $4.1$ and applying topos techniques in $4.2$. We conclude in section $5$.

\section{An Observer's Worldviews}\label{sec:obs}

In this section we will explore how an observer and its worldviews might be defined in terms of standard GR structures. Explicitly relating worldviews with `physical fields' on the spacetime enables us to model not only the observer itself, but how one observer might view another, and how two `worldviews' describing the same spacetime feature might be related.

\subsection{Formalising the Observer}\label{subsec:obs-obs}

In this section we will discuss how an observer might be represented within GR, following standard practise where appropriate. This simple yet critical first step is relatively straightforward within our classical framework, though it would be much less clear how we might proceed if we generalised to a quantum context.

\subsubsection{Generalised Spacetime}\label{subsec:spacetime}

We begin with a spacetime $\Mg$ where $\M$ is a 4D manifold and $g$ a $(0,2)$ non-degenerate metric tensor field of signature $(3,1)$ on $\M$. We will begin by requiring strong causality, and defer a more careful examination of the most appropriate causality condition to further research. For simplicity, unless explicitly mentioned otherwise, we will assume that the manifold can be covered by a single coordinate map.

We can generalise our spacetime by introducing a set $\F$ of tensor fields on $\M$, and will require $\F$ to contain the metric $g$; we can then write $\Mf$ to denote our generalised spacetime. Intuitively we interpret the fields in $\F$ as `physical fields' representing the observables and physical forces in our theory, for example the standard fields used in electromagnetics, or a $\Z_2$ valued field identifying a particular worldline. In a `geometry is all' approach we would only require the metric. Intuitively we require $\F$ to be `comprehensive', describing all physical elements under discussion.

More precisely, we will allow the elements of $\F$ to be either real tensor fields or $\Z_2$ or $\Z$ valued maps on $\M$, or on `regions' within $\M$\footnote{We may wish to restrict the subsets of $\M$ on which we can define our field by restricting the subsets which can be considered `regions'.}. We can think of the tensor fields as representing `physical fields', and the $\Z_2$ and $\Z$ valued maps as representing the presence of objects (such as an observer), a choice from a range of options or other miscellaneous information. We sill say that two such fields $\phi, \psi$ are \emph{similar} if they have the same range, domain and type (in the case of our real tensor fields), and write $\psi \sim\phi$. For simplicity, unless explicitly mentioned otherwise, we will assume that the range of all our fields is the whole of $\M$.

\subsubsection{Introducing the Observer}

Intuitively we think of an observer as a physical system identified with a world tube (or world tube segment) which contains `records' of its past observations, which we might identify with some subset of its past internal states. We require that records, once made, are preserved for the duration of the observer (`no records are lost'). We might extend this intuition to allow the observer to make predictions, or possess a `language' and a logic with which to make statements and construct theories. We would seek to identify all of these elements with some property of the observer's internal state. However for the purposes of this investigation we will not need to enquire into the internal structure of an observer, and as is standard in GR will instead work with an `idealised' observer $\O$ which we identify with a world line $\go$. This world line inherits a total order from the (strongly causal) spacetime $\Mf$. We associate a basis frame field $\mathbf{e} = \{e_{\a}\}$ on the worldline $\go$ with the observer, such that the field is invariant under parallel translation along $\go$. We will further associate with $\O$ a set of coordinate maps $\phi_p$ for $p\in\go$, where each $\phi_p$ is compatible with $\mathbf{e}_p$ (for example $\phi_p$ may be the exponential map generated by $\mathbf{e}_p$). As noted above, unless explicitly mentioned otherwise, we will in what follows assume that $\M$ can be covered by a single coordinate map $\phi$, and will further assume that $\phi_p=\phi$ $\forall p$.

Since we require $\F$ to be comprehensive it must contain a field describing the worldline $\go$. We will write $\go^*:\M\longrightarrow \Z_2$ to denote the $\Z_2$ valued indicator function on $\M$ which represents the worldline of the observer $\O$. We then have:
\[
\go^*(p) = 
\begin{cases}
1 & p\in\go \\
0 & p\not\in\go
\end{cases}
\]

\subsubsection{Observations \& Records}

We will write $\Ko_p$ to denote the `knowledge' or `records' of the observer $\O$ at the point $p$. Intuitively $\Ko_p$ represents all of the observations made by $\O$ as it travels along its worldline up to (and including) the point $p$, and is thus concerned with $\lp$. Since we have required $\F$ to describe `all physical elements under discussion' the observer's knowledge necessarily concerns the fields in $\F$, and factoring in our previous comments we restrict this to $\F |_{\lp}$ (which we define in the obvious manner as $\{\phi|_{\lp}|\phi\in\F\}$). In practise we would expect the observer to have only a partial knowledge of the the field configurations in $\lp$, which might be expressed in terms of probabilities over a space of possibilities. We will assume that `no records are lost', so that:
\beq
\Ko_p \Rightarrow \Ko_q
\eeq 
for $p,q\in\go$ whenever $p\geq q$. 

However a full exploration of this question is beyond the scope of this discussion, and we will for now simply assume that $\Ko_p$ consists of a full specification of all the physical fields on the causal past of $p$,
\beq\label{eq:as-knowledge}
\Ko_p = \F |_{\lp}
\eeq

\subsection{Formalising the `Worldviews'}\label{subsec:obs-dyn}

In this section we discuss the construction of `worldviews' by an observer, explicitly relating them to fields $\F$ on our generalised spacetime. We will complement the knowledge $\Ko_p$ of the observer $\O$ at $p$ with a `probability theory' $\Ptho_p$ (see section \ref{subsec:prob-th}) describing the observer's attempt to `predict' the nature of the generalised spacetime beyond the extent of its knowledge (ie outside $\lp$)\footnote{However in a more sophisticated framework the observer may not have full knowledge of the field configurations in its causal past, and would have to extend its probability theories into this region as well.}. Strictly, the term `prediction' should only refer to an observer's statements regarding its possible future knowledge, however we will use the term more loosely here. Finally, we note that properly we should define the `worldview' of the observer at $p$ to be the combination of $\Ko_p$ and $\Ptho_p$. However to facilitate a comparison with CH we will use the term `worldview' to refer only to $\Ptho_p$ and regard $\Ko_p$ as a feature of the observer itself.

\subsubsection{Stochastic Theories}

In what follows we will adopt a standard probability framework to describe the observer's predictions at a point $p$, though this may of course be generalised. Noting that we are discussing the \emph{observer's} predictive framework rather than the `actual dynamics', we identify three ways in which we expect such a framework to arise:
\begin{enumerate}
\item\label{case:prob-stoch} The observer is using a stochastic theory, and so assumes inherent non determinism in the dynamics.
\item\label{case:prob-past} The observer is using a deterministic theory, but accounts for its lack of knowledge of field configurations in regions spacelike to $\lp$ by introducing a stochastic framework. 
\item\label{case:prob-microstates} The observer is using a deterministic theory, but accounts for its lack of knowledge inside and outside $\lp$ by introducing a stochastic framework.
\end{enumerate}
In case \ref{case:prob-microstates} we might imagine that the observer is unaware of (unable to observe) the fine grained `micro-states' of fields within $\lp$, yet expects these states to have an influence on its future observations. Although this may may seem more realistic than case \ref{case:prob-past}, it contradicts our simplifying assumption that the observer at $p$ has full knowledge of all field configurations within $\lp$. In order to maintain this simplification we will adopt case \ref{case:prob-stoch} and allow the observer to adopt a genuinely stochastic theory, though of course this does still allow for a deterministic dynamics which the observer is unaware of. We leave an exploration of case \ref{case:prob-microstates} to a more detailed analysis.

\subsubsection{Probability Theories}\label{subsec:prob-th}

We now make a predictive framework more formal. At each $p\in\go$ we associate with the observer $\O$ a \emph{probability theory} $\Ptho = \Pthop$ consisting of a \emph{sample space} $\Wo_p$, and \emph{event algebra} $\Lo_p\subseteq 2^{\Wo_p}$, which we require to be a $\s$-algebra, and a \emph{probability measure} $\Po_p$.

\paragraph{The Sample Space \newline}
Now we have written $\F$ to denote the set of physical fields on $\M$, which we have assumed to be tensor fields including the metric. Further, we have assumed that the observer $\O$ at $p$ has full knowledge of the fields specifications $\F |_{\lp}$ in its causal past $\lp$. We then assume that the observer builds on this knowledge to identify the codomain, the tensor type and a set of constraints regarding each field in $\F |_{\lp}$. This allows the observer to extend each field in $\F |_{\lp}$ to the whole space $\M$ in a manner consistent with these constraints. The set of all such extensions of $\F |_{\lp}$ forms the sample space $\Wo_p$.

More concretely, if $R\subseteq\M$ is a spacetime region\footnote{We may of course wish to restrict the subsets of $\M$ which count as regions', but this is beyond the scope of our current discussion.}, we will say that a field $\psi$ is an \emph{extension} of a field $\phi$ over $R$ if $\psi$ is similar (see section \ref{subsec:spacetime}) to $\phi$ everywhere and equal to $\phi$ on $R$,
\begin{eqnarray*}
\psi &\sim& \phi \\
\psi |_R &=& \phi |_R
\end{eqnarray*}
Further, we will say that a set of fields $\w$ is an extension of $\F$ over $R$ if the two sets are of the same cardinality and every element of $\w$ is an extension of a distinct element of $\F$, so that $\w |_R = \F |_R$. We then define $\Wo_p$ as the set of extensions of $\F$ over $\lp$. Note that this is independent of the observer $\O$, which is a consequence of our simplifying assumption that the observer has full knowledge of the field configurations in its causal past. 

\paragraph{Coordinate Systems \newline}
Before continuing we note that in general the observer at $p$ is unaware of the future path of its worldline, and of the relation between the various coordinate systems $\phi_q$ for $q\geq p$. The maps between these coordinate systems would themselves be subject to prediction, and in extending $\F\ |_{\lp}$ it would be more accurate to parameterise $\go$ by proper time and to regard the physical fields as maps with domain $\R^4$ (the images of the coordinate maps), so that the sample space at $p$ would be composed of objects that can be defined by the observer at $p$. However, since we have assumed the existence single coordinate system covering $\M$ we will leave the investigation of these questions to future research, and continue to use fields with domain $\M$ in our samples spaces.

\paragraph{The Event Algebra \newline}
The event algebra $\Lo_p\subseteq 2^{\Wo_p}$ consists of the subsets of $\Wo_p$ which we would like the observer at $p$ to treat as dynamical (and logical) predicates. We will insist that $\Lo_p$ is a Boolean algebra, and in general probability theories it is typically a $\s$-algebra generated by measurable subsets of the sample space.

Intuitively, we would like our `dynamical predicates' to be `localisations' associated with particular regions of spacetime. To this end, given a field $\psi\in\F$ we want to write $\E_p(\psi(R))$ to denote the event (in $\Lo_p$) that $\psi$ takes a particular value in the region $R\in\M$ (which might be a single point). More precisely, we define $\E_p(\psi,\Psi, R)$ to the set of all $\w\in\Wo_p$ in which the extension of the field $\psi |_{\lp}$ takes the values $\Psi$ in the region $R$ for some field $\Psi$ which is similar to $\psi$ on $R$. We will say that $R$ is the \emph{support} of the event $\E_p(\psi,\Psi,R)$, and write $R=\Rc(\E_p(\psi,\Psi,R))$. We might then take $\Lo_p$ to be the $\s$-algebra generated by events which can be expressed in the form $\E_p(\psi,\Psi,R)$. However we will leave this detailed discussion to future research, and for now will simply take $\Lo_p = 2^{\Wo_p}$ and assume that every element is measurable. 

To further develop our notation, we will write $\E_p(\psi(R))$ as a shorthand for $\E(\psi,\Psi,R)$ in the case that we do not need to define the actual function $\Psi$. We will further write $\E_p(\F(R))$ to denote $\bigcap_{\psi\in\F}\E(\psi(R))$.

\paragraph{The Probability Measure\newline}
We will require $\Po_p$ to be a probability measure satisfying the usual axioms:
\begin{enumerate}
\item $\Po_p:\Lo_p\rightarrow [0,1]$
\item $\Po_p(\Wo_p) = 1$
\item $\Po_p(A\sqcup B) = \Po_p(A) + \Po_p(B)$
\end{enumerate}
where $A,B\in\L_p$ are disjoint, $A\cap B=\emptyset$, and $\sqcup$ denotes disjoint union.

We will not proscribe the measure, but allow our observers to construct dynamical theories themselves. Note that the measures derived by the observers will not necessarily align with any `true dynamics' which is available to us, nor will the theories of two observers necessarily be consistent with each other.

\subsubsection{Causal Consistency}\label{subsec:prob-theory-causality}

We now consider how the inter-relationships between the probability theories of a particular observer $\O$ correspond to the  spacetime causal structure, beginning with the sample space and event algebras. 

\paragraph{The Sample Space and Event Algebras \newline}
Consider points $p,q\in\go$ such that $p\leq q$. Now by construction every $\w\in\Wo_q$ agrees with $\F$ on $J^-(q)$, so noting that
\[
p\leq q\Rightarrow \lp\subseteq J^-(q)
\] 
we see that every $\w\in\Wo_q$ agrees with $\F$ on $\lp$, so that:
\beq
p\leq q\Rightarrow\Wo_p\supseteq\Wo_q
\eeq
Since we have taken $\Lo_p=2^{\Wo_p}$, this implies a similar result for our event algebras, 
\beq\label{eq:event-algebra-causality}
p\leq q \Rightarrow \Lo_p\supseteq\Lo_q
\eeq
Notice that we can reformulate this as a descending filtration by defining $\Wo=\bigcup_{p\in\go}\Wo_p$ and considering the $\L_p$ as subalgebras of $2^{\Wo}$ in the obvious manner. Now (\ref{eq:event-algebra-causality}) turns out to be central in our attempts to construct a global logical structure (section \ref{subsec:obs-logic}), but we have derived it from the naive choice of $\Lo_p=2^{\Wo_p}$. If we were to generalise to a more sophisticated formulation of event algebras, we might impose (\ref{eq:event-algebra-causality}) as a constraint.

\paragraph{The Probability Measures \newline}
Now we have not defined the measure, but have left it for the observers to construct. We are then unable to infer consistency conditions from the causal structure, but will instead impose them as constraints upon the theories we are allowing our observers to build. We will impose three consistency conditions:

\begin{enumerate}
\item
If two events $A,B\in\Lo_p$ have spacelike supports then we will require them to be independent,
\beq\label{eq:spacelike-independence}
\Rc(A),\Rc(B) \, \, \text{spacelike} \, \Rightarrow \Po_p(AB)=\Po_p(A)\Po_p(B)
\eeq 

\item 
We might expect the probability assigned by an observer to a future event to alter as the observer gains more information. However we will require that these probabilities should be consistent in that if the observer correctly predicts information it will gain in the future, then the probability it assigns to an event $A\in\Lo_q\subseteq\Lo_p$ is the same as the probability assigned at $q$:
\beq\label{eq:consistency}
\Po_q(A) = \Po_p(A | \E_p(\F |_{J^-(q)}))
\eeq

\item
Our third condition is intuitively similar to a combination of the previous two, and allows us to create independence by conditioning on joint past of an event. Consider two events $A,B\in\Lo_p$, with supports $\Rc(A),\Rc(B)\in\M$. The causal structure identifies the region $J^-(\Rc(A))\cap J^-(\Rc(B)$ as responsible for any joint behaviour in $\Rc(A),\Rc(B)$, so if we introduce the event,
\[
P=\E_p(J^-(\Rc(A))\cap J^-(\Rc(B))
\]
we can then impose:
 \beq\label{eq:conditional-independence}
\Po_p(AB \, | \, P) = \Po_p(A \, | \, P) \, \Po_p(B \, | \, P)
\eeq 

\end{enumerate}

\subsection{Comparison with Consistent Histories}\label{subsec:ch-comparison}

The structures we have defined are directly comparable to consistent histories, with the probability theories of our observers' worldviews acting as the equivalent for the probability theories of the consistent set worldviews. However, CH employs multiple worldviews only because they are (in general) mutually incompatible so that a single worldview is insufficient. Moreover, this multiplicity of worldviews is held to be the defining feature of `quantum' as opposed to `classical' behaviour. It is then natural to ask whether our worldviews are in some way mutually incompatible.

Now in one sense our probability theories are trivially compatible, in that they are all derived from a single generalised spacetime. However this misses the point, as the spacetime acts as a model of an `objective reality' which is not fully available to our observers. Further, the way in which the observers construct their worldviews is itself a feature of the spacetime, and might interact with the features under observation. It turns out that we can in fact find `incompatibilities' between various worldviews, for much the same reason that this occurs in QM; the details of the measurement process.

\subsubsection{The Measurement Problem in General Relativity}\label{subsec:measurement}

We will focus on a pair of interacting observers which are both observing the configuration of the same field in the same region of spacetime. We ask whether their two worldviews are intuitively `compatible' in a manner analogous to the use of the term in CH. 

\begin{figure}[!htb]
  \centering
   \includegraphics[scale=0.8]{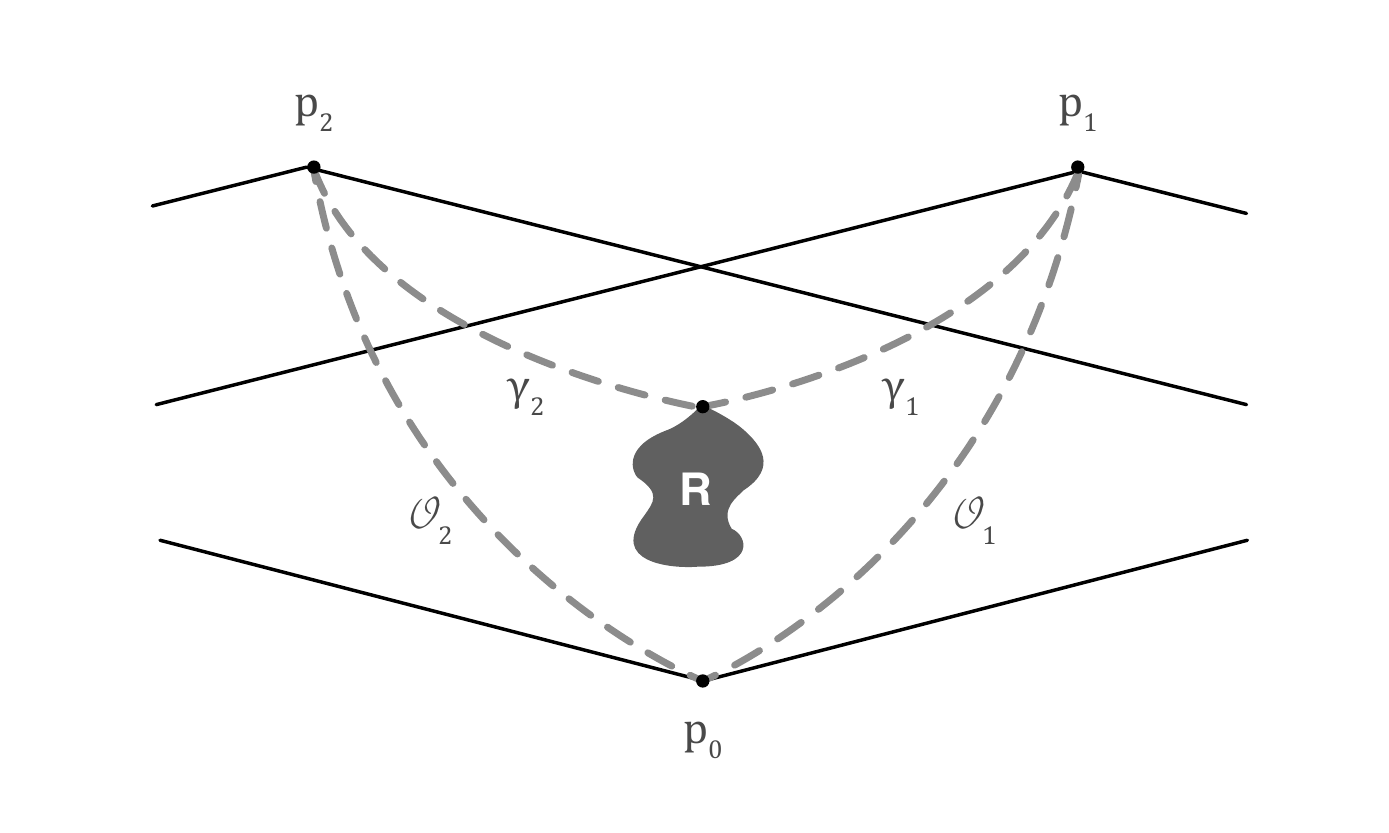}
  \caption{The Measurement Problem}\label{fig:measurement}
\end{figure}

To make this more precise, consider an extended spacetime $(\M,\F)$ containing two observers $\O_1, \O_2$. The observers meet at $p_0\in\M$ and then travel to the spacetime points $p_1,p_2$ respectively, where  they observe the configuration of the field $\psi\in\F$ in the region $R\subset J^+(p_0)\cap J^-(p_1) \cap J^-(p_2)$. More accurately signals travel from $R$ to $p_1$ and $p_2$ along paths $\g_1,\g_2$, carrying information regarding $\psi(R)$. These signals are then measured by the two observers at $p_1$ and $p_2$ respectively (see figure \ref{fig:measurement})

Now we can set these measurements to have binary `yes/no' outcomes, which we represent by two $\Z_2$ valued fields $\phi_1, \phi_2\in\F$ respectively\footnote{It may be more tidy to restrict the ranges of the fields $\phi_1,\phi_2$ to the points $p_1,p_2$ respectively.}. We can then define the events $E_1^+ = \E^{{\O_1}}_{p_0}(\phi_1,1,p_1),E_1^- = \E^{{\O_1}}_{p_0}(\phi_1,0,p_1)$ in the event algebras $\L^{\O_1}_{p_0}$ to denote the measurement at $p_1$ resulting in a `yes' ($E^+$) or a `no' ($E^-$). We can similarly define $E^+_2,E^-_2$ in $\L^{\O_2}_{p_0}$. 

Now there is no reason that the observers can not make predictions regarding each others future measurements, so we could just as well have defined $E_1^+,E_1^-$ in $\L^{\O_2}_{p_0}$, or $E_2^+,E_2^-$ in $\L^{\O_1}_{p_0}$. In fact, as we shall see in section \ref{sec:global}, our strong assumptions mean that the two event algebras can be identified. For simplicity we will then, without loss of generality, work in $\L^{\O_1}_{p_0}$, and assume that all our events are in this algebra. Note that $\pi_1=\{E_1^+,E_1^-\}$ and $\pi_2=\{E_2^+,E_2^-\}$ are then two partitions of $\W^{\O_1}_{p_0}$.

Now let us imagine that the two observers are attempting to make identical measurements of $\psi(R)$, so we might think of the two partitions $\pi_1,\pi_2$ as two `views' of the same physical phenomenon. We would then say that the two views are compatible if they are equivalent, such that $E_1^+=E_2^+$ and  $E_1^-=E_2^-$. We would then be able to ignore the details of each observer and work only with the field under observation.

However even a cursory look at the problem shows us that in general this is not the case; the details of the measurements can not be ignored.  For example, the measurements at $p_1$ and $p_2$ might involve a comparison of the information received from $R$ with fields $\chi_1,\chi_2$ representing measurement devices `carried' by the respective observers. Even if $\chi_1(p_0)=\chi_2(p_0)$ we are not guaranteed that the two measurements will be always yield equivalent results. In particular, the values of these `measurement fields' at $p_1$ and $p_2$ may become correlated with each other, with $\psi$  or with the measurement process itself, adding to the complication.

This leaves us with two incompatible worldviews $\pi_1,\pi_2$. However, this `incompatibility' seems weaker than the incompatibility of consistent sets in CH, for our four events lie within the same Boolean algebra and may simultaneously receive truth values. To test whether this is the case, we will in the next section adapt the CHSH framework to try to isolate the gap between incompatibility in CH and GR.

\section{CHSH in Curved Spacetime}\label{sec:chsh}

In section \ref{subsec:measurement} we saw that interacting observers could have incompatible worldviews, though this incompatibility seemed weaker than that which is found in CH. In this section we will build on these ideas to explore the gap between our observer dependent formulation of GR and the CH formulation of QM. As discussed in section $1$ it is hoped that this will allow us to isolate and better understand the truly `quantum' as opposed to the simply `observer dependent' behaviour. To achieve this we turn to the EPR-Bell-CHSH argument which was developed precisely to identify and test this gap, and was experimentally verified by Aspect \cite{aspect}.

The discussion began with the EPR paper of 1935 \cite{epr} in which Einstein, Podolsky and Rosen argued that Quantum Mechanics could not be considered as what they termed a `complete' theory of `objective reality'. The EPR argument was developed further in 1964 by Bell \cite{bell:1964}, who argued that `quantum statistics' are in contradiction with the combination of `causality and locality', with the well known Bell inequality providing a means of distinguishing quantum from classical behaviour. In 1969 Clauser, Horne, Shimony and Holt developed Bell's argument even further \cite{chsh}, providing another inequality (the CHSH inequality) with which to distinguish quantum from classical behaviour, and a thought experiment through which their inequality might be tested. Finally, this thought experiment was implemented as an actual experiment by Aspect, Dalibard and Roger \cite{aspect}, who found `non-classical' quantum behaviour to occur. This argument is core to much of our understanding of quantum as opposed to classical behaviour, and is discussed at length by many authors, see for example Bell's discussion in \cite{bell:speakable} or the account given in \cite{aharonov:paradoxes}.

We note that in all cases the argument is formulated in a flat spacetime, and it is generally assumed that the presence of curvature will not have a material effect. This seems odd, given the critical role of comparing the directions of measurement at different spacetime points, and we will see that curvature does in fact play a material role.

We will proceed cautiously so as to properly account for the subtle effects of curvature. In section \ref{subsec:chsh-setup} we describe a CHSH framework in a generalised spacetime, including a careful account of the observers and how the make their measurements. In section \ref{subsec:chsh-dyn} we discuss the dynamics, and contrast our results with the standard argument on a flat spacetime.  In section \ref{subsec:chsh-discussion} we round up and discuss the relevance of our findings to the discussion of observer dependence.

\subsection{The CHSH Construction}\label{subsec:chsh-setup}

\subsubsection{Overview}

\begin{figure}[!htb]
  \centering
   \includegraphics[scale=0.8]{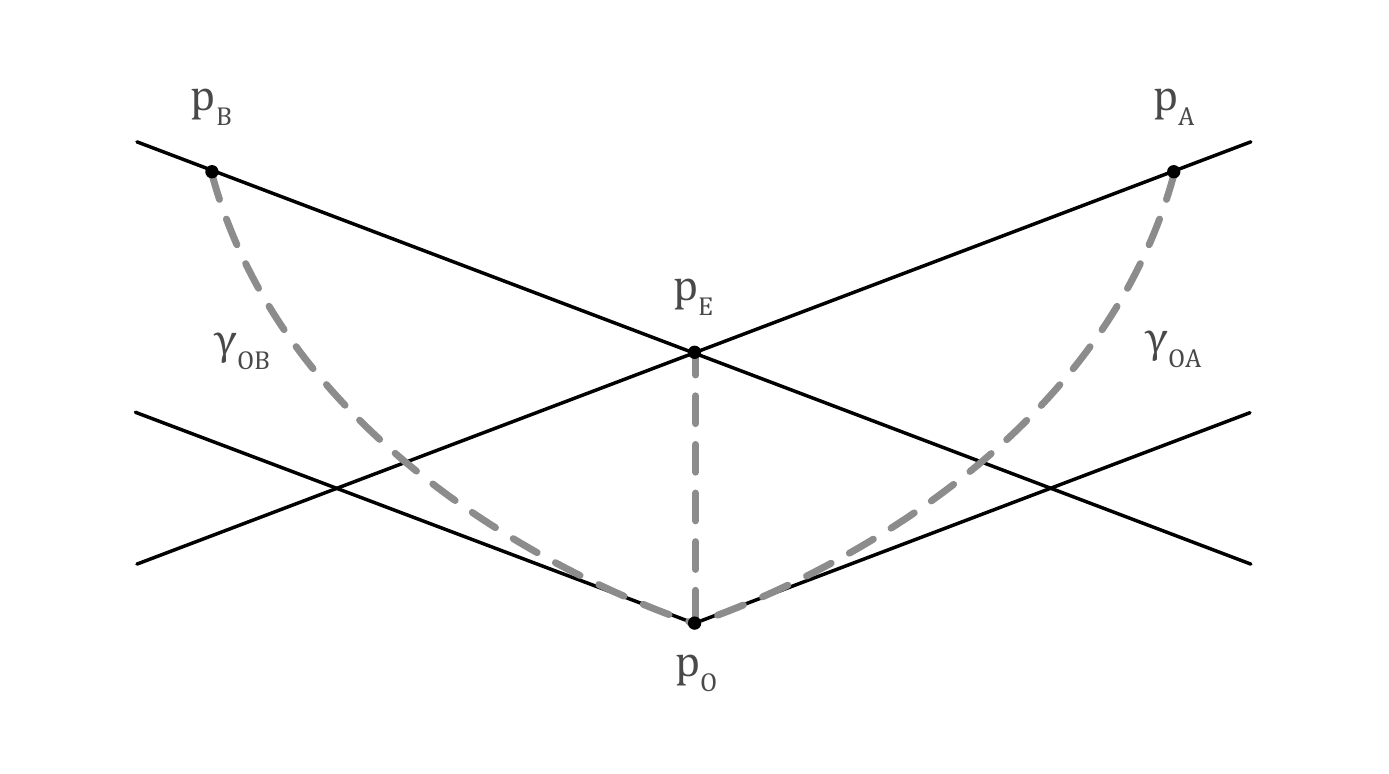}
  \caption{The CHSH thought experiment}\label{fig:chsh}
\end{figure}

Two observers $\Oa$, $\Ob$  meet at a spacetime point $p_O\in\M$ to set up the CHSH experiment; we think of them as making predictions at $p_O$ and then testing these predictions by carrying out the experiment. The observers synchronise their coordinate systems, set up the experimental apparatus $\Pi$ and agree on two sets of measurement directions (one for each observer). The apparatus $\Pi$ travels to $p_E\in\M$ where it emits two beams of light, of random and opposite linear polarisation. The two beams are intercepted by the observers at spacelike points $p_A$ and $p_B$ respectively, where their polarisations are measured in directions which are randomly chosen from the options agreed on at $p_O$. The choices of measurement direction are regarded as experimental `inputs'. The two observers then meet again at a final point $p_F\in\M$, where they can `compare notes' and combine their observations. The observers collaborate at $p_O$ so that their probability theories are equivalent, and contain the predictions they make concerning the outcomes to be observed at the points $p_A$ and $p_B$. The experiment is outlined in Figure \ref{fig:chsh}, with the final point $p_F$ omitted. 

\subsubsection{Notation}

As before we are assuming a generalised spacetime $\Mf$, and will in general denote points in $\M$ by $p_X$ for some $X$.  We will write $\g_{XY}$ to denote the appropriate spacetime geodesics connecting points $p_X$, $p_Y\in\M$, and we will write $\G_{XY}$ to denote parallel transport from $p_X$ to $p_Y$ along $\g_{XY}$. In general there will only be a single relevant path connecting any two of the points we are interested in, which will be clear from the context. We write $u_X$ to denote a vector in $T_{p_X}(\M)$, and where the meaning is unambiguous we will write $u_Y$ to denote its transport $\G_{XY}(u_X)$ to $p_Y$ along $\g_{XY}$. Finally, where appropriate we will use the subscripts $A$, $B$ for objects associated with either of the two observers $\Oa$, $\Ob$. 

Due to the complexity of the problem and the number of variables involved we will refrain from explicitly expressing each object in terms of a field in $\F$ so that we maintain readability. Were we to properly describe each field our account would become unnecessarily cluttered and opaque. We will instead simply refer to vectors and variables `in' or on $\M$, and assume that these can be expressed appropriately in terms of elements of $\F$.

\subsubsection{The initial point, $p_O$}

At $p_O$ the observers will perform the initial setup of the experiment, which we are idealising to occur at a single point.
This set up includes agreeing several key directions (ie tangent vectors) which will be important during the experiment. 

Firstly, the observers synchronise their basis frames, by agreeing on an orthonormal $1\times 3$ basis frame $e_O  = \{e_{O\a}\}_{\a=0}^3$ at $p_O$. As is standard, we assume that $e_{O0}$ is timelike while $e_{Oi}$ is spacelike for $i\in\{1,2,3\}$. The observers will then `carry' this basis frame with them by translating it along their respective worldlines to use throughout he experiment. We further assume that $e_O$ is compatible with some coordinate system $\c_O$ around $p_O$.

Next, the two observers set up the the experimental apparatus $\Pi$, including setting the spacelike unit vectors $\da_O$, $\db_O$ which the apparatus will use to determine the directions in which to emit its polarised light beams at $p_E$. We will assume that $\db_O=-\da_O$. For simplicity we abbreviate $d_O = \da_O$, so that $\db_O = -d_O$. Without loss of generality we will assume that $d_O = e_{O3}$. The proper time $\t_E$ which the apparatus will wait before emitting the beams is also set at $p_O$. Notice that we should properly be using a field in $\F$ to describe these vectors. However as mentioned above, for simplicity we will not explicitly refer to fields except where necessary for clarity.

Finally, the observers will set two spacelike 2D \emph{measurement planes}, $\ma_O,\mb_O$ in the tangent plane $T_{P_O}(\M)$, which are defined by being orthogonal to the directions $\da_O, \db_O$ respectively. Then since we have assumed $\db_O=-\da_O=e_{O3}$ the two planes will be identical and spanned by $\{e_{O1},e_{O2}\}$; we will abbreviate to $m_O = \ma_O = \mb_O$. Within each measurement plane the observers define a set of potential measurement directions, $\{a_O^i\}_{i\in I_A}\in m_O$ associated with $\Oa$ and $\{b_O^j\}_{j\in I_B}\in m_O$ associated with $\Ob$, where the $a_O^i,b_O^j$ are unit vectors. The indexing sets $I_A,I_B$ may be uncountable, however for now we will assume they are both finite. In fact, actual realisations of this experiment have used two possibilities for each observer \cite{aspect}.

\subsubsection{The path of the apparatus, $\g_{OE}$} 

The apparatus $\Pi$ travels from $p_O$ to $p_E$ along $\g_{OE}$, which we therefore take to be timelike. For simplicity we will assume that the experiment is set up so that $\g_{OE}$ is a timelike geodesic. In particular we will assume that the observers at $p_O$ set its initial tangent vector $\dot{\g}_{OE p_O}$ to equal $e_{O0}$ so that the apparatus appears to be stationary in $\c_O$. We know that $\g_{OE}$ has length $\t_E$, which is the proper time set at $p_O$ that the apparatus `waits' before emitting its light beams.

\subsubsection{The point of emission, $p_E$}

The apparatus $\Pi$ emits two light beams in the directions $d_E, -d_E$ where $d_E = \G_{OE}(d_O)$ is the parallel translation of $d_O$ from $p_O$. The light beams are linearly polarised in the directions  of the unit spacelike vectors $\va_E, \vb_E$  respectively. We require that the polarisations are in opposite directions, $\vb_E = -\va_E$, allowing us to abbreviate by introducing $v_E$ and dropping the $A, B$ superscripts, $\va_E = v_E$, $\vb_E = -v_E$.

Note that the polarisation directions are spacelike and must be orthogonal to the direction of the beam, so that by construction they will lie within the measurement plane $m_E = \G_{OE}(m_O)$ parallel translated from $p_O$. We can then define the $\thv$ to be the angle in $m_E$ between $v_E$ and $e_{E1}=\G_{OE}(e_{O1})$. We will assume that this angle is `random' in that it is independent of all other variables in our experiment, other than the measurement outcomes $A$ and $B$ defined below.

\subsubsection{The paths of the observers, $\g_{OA}$ and $\g_{OB}$}

The observers $\Oa$,$\Ob$ travel from $p_O$ to $p_A$,$p_B$ along the worldline segments $\g_{OA}$,$\g_{OB}$ respectively. For simplicity we will focus on $\Oa$, but our discussion will be equally valid for $\Ob$.

The observer $\Oa$ positions itself so as to intercept the signal that it has set up to be emitted form $p_E$. To achieve this, and noting that the apparatus will be stationary in the observer's coordinate system at $p_O$, it will `set out' in the spacelike direction $d_O$ in which it has instructed the apparatus to emit the signal. Concretely, this means that the projection of the initial tangent vector $\dot{\g}_{OA p_O}$ on the subspace spanned by the spacelike basis vectors $\{e_{Oi}\}_{i=1}^3$ will be a positive multiple of $d_O$. For $\Ob$ the initial tangent vector will be in the $-d_O$ direction.

\begin{figure}[!htb]
  \centering
  \includegraphics[scale=0.8]{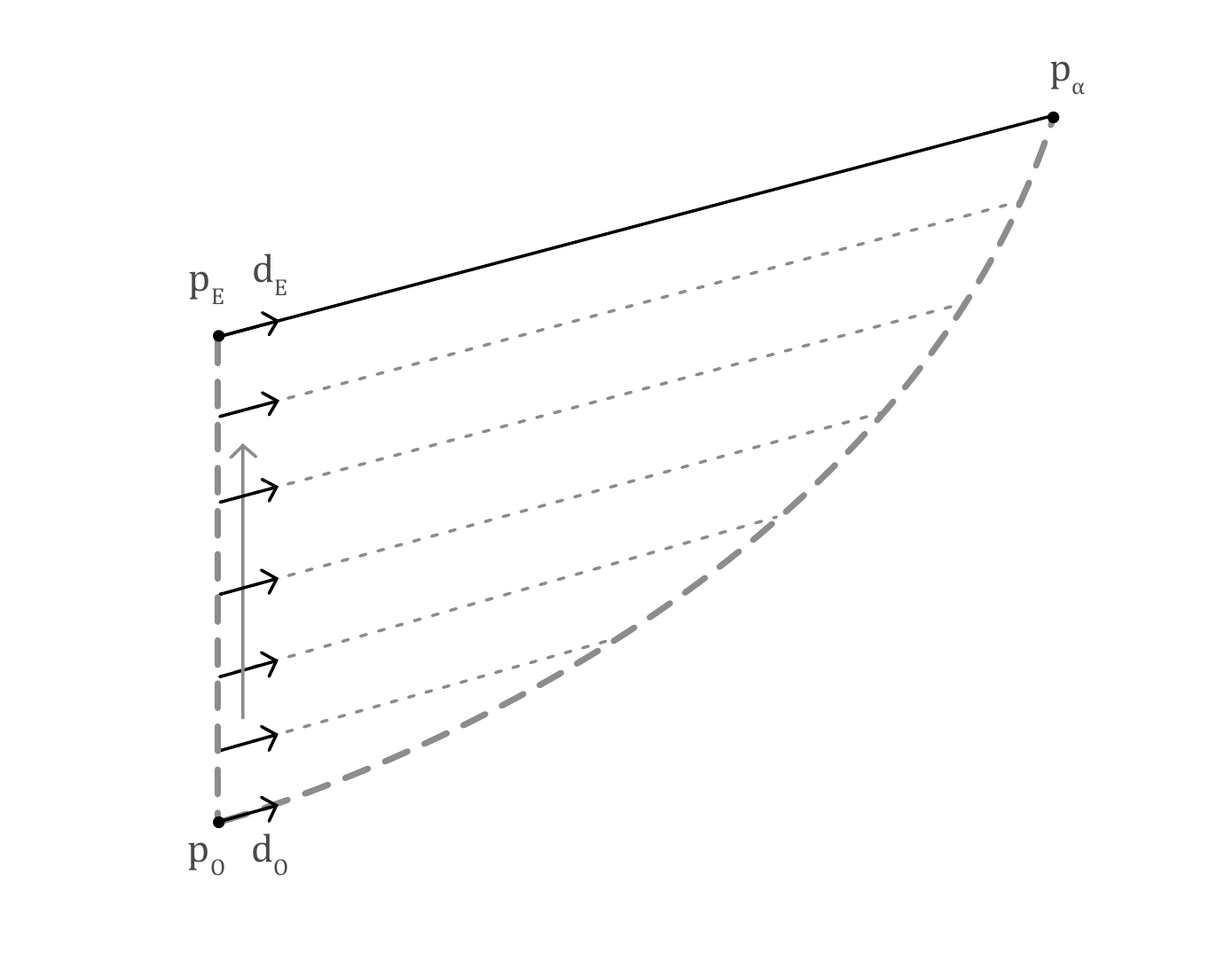}
  \caption{The Observers Path}\label{fig:observer-paths}
\end{figure}

Now at $p_O$ the observer $\Oa$ is unaware of the nature of the spacetime regions it and the apparatus will pass through, in particular it is not aware of the curvature. This means it may have to alter or correct its planned course so as to be in place to intercept the emission from $p_E$. To achieve this, we can imagine the apparatus emitting a continuous signal in all directions as it progresses along $\g_{OE}$, this signal should contain direction information which the observer may use to correct its path (see figure \ref{fig:observer-paths}).

\subsubsection{The paths of the polarised emissions, $\g_{EA}$ and $\g_{EB}$}

The polarised beams emitted at $p_E$ travel along the lightlike geodesics $\g_{EA}, \g_{EB}$ to reach the observers at $p_A, p_B$. As previously discussed, the apparatus $\Pi$ is set up so that these beams are emitted in the $d_E$ and $-d_E$ directions respectively. Concretely, this means that if $\dot{\g}_{EA p_E}$ is the tangent vector to $\g_{EA}$ at $p_E$, its projection onto the spacelike subspace spanned by $\{e_{Ei}\}_{i=1}^3$ is a positive multiple of $d_E$. Similarly the projection of $\dot{\g}_{EB p_E}$ is $-d_E$.

\subsubsection{The points of measurement, $p_A$ and $p_B$}\label{subsec:measurement-points}

At $p_O$ the observer $\Oa$ has with it the parallel transport $m_A$ of the measurement plane $m_O$ along $\g_{OA}$. Further, $\Oa$ has the transported set of potential measurement directions $\{a_A^i\}_{i\in I_A}$ where $a_A^i = \G_{OA}(a_O^i)$. We can similarly define $m_B$ and $\{b_B^i\}_{i'in I_B}$. Further, $\Oa$ receives the beam $\g_{EA}$ from $p_E$, included the transported polarisation vector $v_A = \G_{EA}(v_E)$. Similarly $\Ob$ receives $-v_B=\G_{EB}(-v_E)$.

To make the measurement, the observer $\Oa$ `randomly' chooses one of its potential measurement directions, $a^{i_A}_A\in\{a_A^i\}_{i\in I_A}$, by which we will mean that $i_A$ is independent of all variables in our experiment except for the measurement outcome $A$ (see below). Similarly $\Ob$ `randomly' chooses $b^{j_B}_B\in\{b_B^i\}_{i\in I_B}$. Note that the observers are in effect choosing $i_A$ from $I_A$ and $j_B$ from $I_B$. To simplify the notation we will abbreviate $a^{i_A}_A$ to $a_A$ and $b^{j_B}_B$ to $b_B$. Finally, $\Oa$ measures the polarisation of the beam in the $a_A$ direction (and records the result). Similarly $\Ob$ measures the polarisation of its beam in the $b_B$ direction. Note that the choices $i_A,i_B$ of measurement direction are in fact $\Z$ valued fields on $p_A,p_B$.

Now note that the polarisation direction $v_A$ will not in general lie within the measurement plane $m_A$, but rather in the plane $\bar{m}_A = \G_{EA}\G_{OE}(m_O)$. We will write $\phi_A$ to denote the angle between these two planes, and $\th_A(a_A,v_A)$ or simply $\th_A$ to denote the angle between $a_A$ and the projection $\nu_A$ of $v_A$ onto $m_A$. Note that $\th_A$ is preserved by parallel translation toward $p_O$ along $\g_{OA}$, since both $v_A$, $m_A$ and $a_A$ would be translated together. We will also need the angles $\tha$ (in $m_A$) between $a_A$ and $e_{A1}$ and $\thb$ (in $m_B$) between $b_B$ and $e_{B1}$.

\subsubsection{The measurements at $p_A$ and $p_B$}

The measurement outcomes at $p_A$ and $p_B$ are binary, taking values $\pm 1$. Intuitively, we would like a measurement outcome of $+1$ to occur if the polarisation is found to be in the measurement ($a_A$ or $b_B$) direction, with a $-1$ outcome to occur otherwise. This might for example be implemented by passing the light emissions through polarisation filters based on the measurement directions, with a $+1$ outcome being recorded if `sufficient' light passed through. For more details of an actual implementations see \cite{aspect}.

As in section \ref{subsec:measurement} we will represent the measurement outcomes by physical fields in $\F$:
\begin{eqnarray}
A:p_A&\longrightarrow& \{-1,+1\} \nn \\
B:p_B&\longrightarrow& \{-1,+1\}
\end{eqnarray}
As these variables are only defined at a single point, we will abuse notation by also using $A,B$ to refer to the actual outcomes $A(p_A),B(p_B)$ respectively.  While this notation is somewhat confusing, since we are using `A' both as a label, as a variable and as a value, it is standard in accounts of CHSH so we will adopt it.

\subsection{The Dynamics}\label{subsec:chsh-dyn}

As discussed above, the observers collaborate at $p_O$ to align their worldviews and set up the experiment, which test the predictions they make at $p_O$. Now in our framework $\Mf$ acts as the `objective reality' (in the EPR terminology \cite{epr}) in which our observers are operating. While we are able to access the actual outcomes $A(p_A),B(p_B)$, our observers at $p_O$ can not. It is then precisely the predictions made by the observers at $p_O$ (rather than our knowledge of the actual outcomes) which we will be interested in when discussing the dynamics of this experiment.

More concretely we are interested in the probability theories $\Pth^{\O_A}_{p_O},\Pth^{\O_B}_{p_O}$ used by the observers at $p_O$. As in section \ref{subsec:measurement} we assume that the collaboration between the observers is such that we can identify the two theories, $\Pth^{\O_A}_{p_O}=\Pth^{\O_B}_{p_O}$, and without loss of generality assume that we are using $\Pth^{\O_A}_{p_O}$, which we abbreviate to $\Pth_O=(\W_O,\L_O,\P_O)$. We can think of this as a `joint probability theory' derived by the observers during their collaboration at $p_O$.

We can then define the following events in $\Pth_O$ representing the possible measurement outcomes:
\begin{eqnarray*}
A^+ &=& \E^{\O_1}_{p_O}(A,+1,p_A) \nn \\
A^- &=& \E^{\O_1}_{p_O}(A,-1,p_A) \nn \\
B^+ &=& \E^{\O_1}_{p_O}(B,+1,p_B) \nn \\
B^- &=& \E^{\O_1}_{p_O}(B,-1,p_B) \\
\end{eqnarray*}
As in section \ref{subsec:measurement}, $\{A^+,A^-\}$ and $\{B^+,B^-\}$ are two partitions of $\W_O$. We can further abbreviate our notation, writing $\E_O$ in place of $\E^{\O_1}_{p_O}$. We can then define the events $\E_O(A(p_A)), \E_O(B(p_B))$ in $\L_O$, which by abuse of notation we further abbreviate to $A,B$.

Now our choices of measurement direction $i_A,i_B$ (section \ref{subsec:measurement-points}) are $\Z$ valued maps on the measurement points $p_A,p_B$ respectively. We can then consider the corresponding events $\E_O(i_A(p_A)), \E_O(i_B(p_B))$ in $\L_O$, which by abuse of notation we abbreviate to $i_A,i_B$ as before. We will in fact make this abbreviation in general, simplifying our statements by abusing notation to write X in place of $\E_O(X)$ for a field $X$.

We are now in a position to state the goal of this section more formally. We would like to determine the probabilities assigned by our observers at $P_O$ to the events $\E_O(A)\cap\E_O(B)$. Since we regard the choices $i_A,i_B$ of measurement direction as inputs we will condition on this rather than attempting to assign to them probability distributions which we can integrate over. Our goal is then to determine the dynamical expression:
\beq\label{eq:chsh-def-dyn}
\P_O(AB \, | \, i_Ai_B)
\eeq

Note that we are always assuming that the experiment is completed successfully, with the observers reaching $p_A,p_B$ and intercepting the signal from $p_E$ as expected. To be accurate we would introduce an event $S$ that the experiment is successful, which we would then condition on. We would then be looking for $\P_O(AB \, | \, i_Ai_B S)$, where we assume that $S$ is independent of our other variables. Will will however continue to `silently' assume that we have conditioned on $S$ without explicitly including it in our expressions. 

\subsubsection{Decomposing the angle $\th_A$}\label{subsec:th-decomp}

The effect of holonomy on the angle $\th_A$ is central to our argument, for in translating the basis frame defining $a_A$ to $p_A$ via $\g_{OA}$ while translating the basis frame defining $v_A$ via $\g_{EA}\circ\g_{OE}$, we have made our angle dependent on the holonomy around the loop $\g_{OEA}=\g_{OE}\circ\g_{EA}\circ\g_{AO}$. This loop is in the future of $p_O$, and so the holonomy is not known by the observers at $p_O$. The correlations between this holonomy and other variables will play a key part in what follows.

\begin{figure}[!htb]
  \centering
  \includegraphics[scale=0.8]{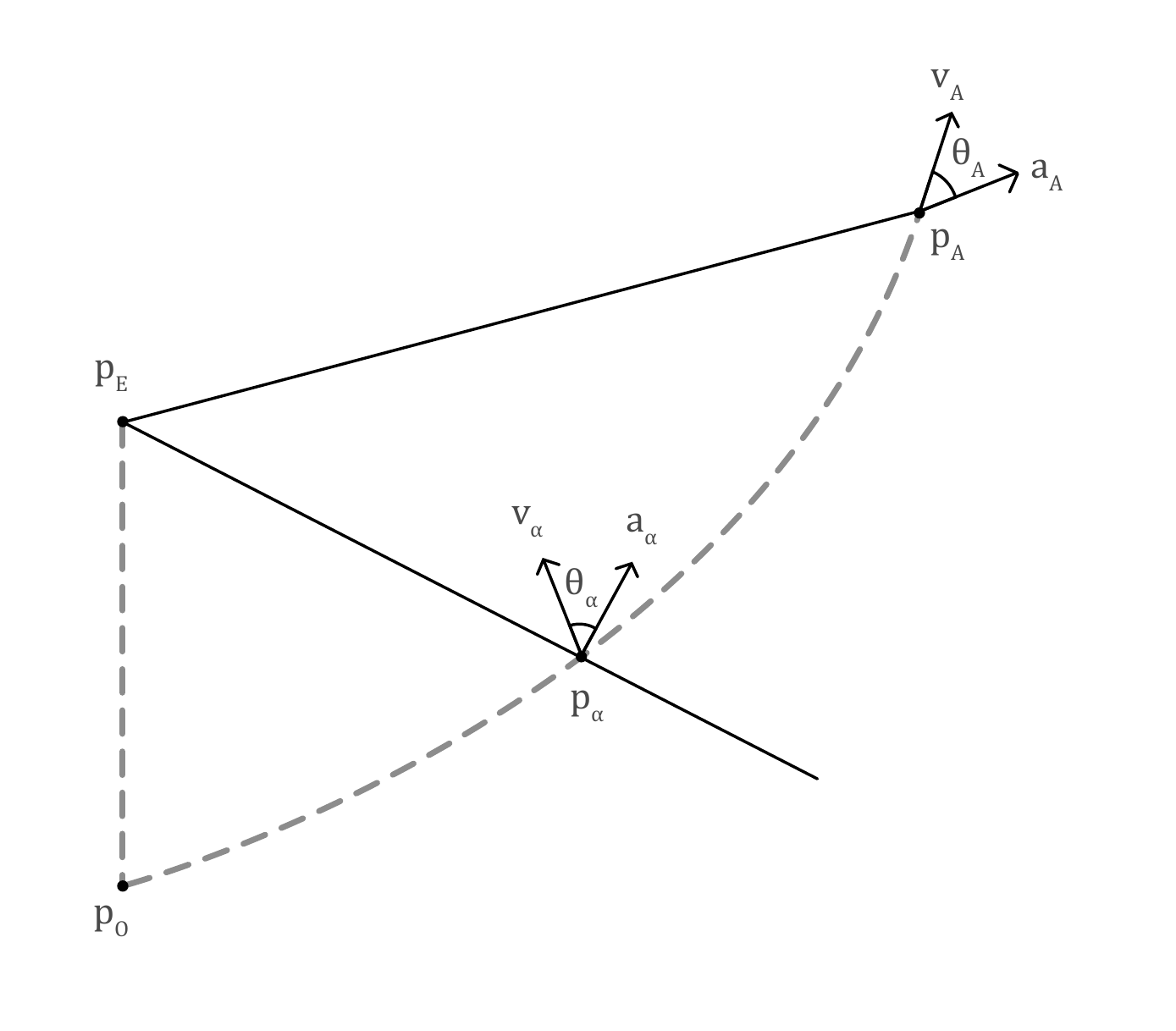}
  \caption{Decomposing the Holonomy}\label{fig:hol-decomp}
\end{figure}

Now although $\g_{OEA}$ is entirely in the future of $p_O$, it is partially in the past of $p_E$. By introducing the point $p_\a$ at the intersection of $\g_{OA}$ and the past lightcone of $p_E$ we can split the loop $\g_{OEA}$ into $\g_{OE\a}$ and $\g_{\a EA}$, with the lightlike geodesic $\g_{\a E}$ forming a common `side' of both loops (see figure \ref{fig:hol-decomp}). Since $\g_{OE\a}$ lies entirely within $\le$ while (with the exception of $\g_{\a E}$) $\g_{\a EA}$ is outside $\le$, we can use this split to isolate the holonomy effects due to factors within $\le$. 

We work through the details of the decomposition in appendix \ref{appendix:thA-thB-decomp}, to yield:
\bea
\th_A &=& \thav + \thao + \thaa \nn \\
\th_B &=& \thbv + \thbo + \thbb \label{eq:thA-thB}
\eea

where $\thao$ is the holonomy around the loop $\g_{OE\a}$ and $\thaa$ the holonomy around the loop $\g_{\a EA}$, with $\thbo$ and $\thbb$ similarly defined (see appendix \ref{appendix:thA-thB-decomp}).

\subsubsection{Conditioning on the Joint Past}\label{subsec:chsh-pe}
 
 As a first step toward determining (\ref{eq:chsh-def-dyn}) we will begin by simplifying our set up by conditioning on the joint past of of the measurement points $p_A,p_B$ and the point of experiment $p_E$. To this end we introduce the event:
 \[
 E = \E_O(J^-(p_A)\cap J^-(p_B))\cap\E_O(j^-(p_E)
 \]
 in $\L_O$. Now $p_E$ is in the joint past of $p_A$ and $p_B$, so we have,
 \[
 E = \E_O(J^-(p_A)\cap J^-(p_B))
 \]
 For convenience, we will write $\P_E$ to denote the measure $\P_O$ conditional on $E$, so that,
 \[
 \P_E(X) = \P_O(X \, | \, E) \, \, \, \forall X\in\L_O
 \]
 It is easy to see that $\P_E$ is a probability measure on $\L_O$.

By conditioning on $E$ we have then `fixed' both the actual results of the experiment and the spacetime region upon which the measurement processes have a joint dependency. Then by our consistency condition (\ref{eq:conditional-independence}) we would expect the resulting measurements to be independent. Our aim in this section is to determine the conditional probability:
\[
\P_E(AB \, | \, i_Ai_B)
\]
 
Now we assume that after making this simplification the outcome $A$ depends only upon the angle $\th_A$ between $a_A$ and the projection of $v_A$ onto $m_A$. We might imagine that the measurement is undertaken using a planar device, so that $\phi_A$ is ignored. There is of course a possibility that $v_A$ will be orthogonal to $m_A$; we will assume that this eventuality is of measure zero and we will ignore it. In practise, depending on the details of the experiment, having $v_A$ orthogonal to $m_A$ might lead to a negative ($-1$) result being recorded as the polarisation will not be found to be in the $a_A$ direction.

As mentioned above, since we have conditioned on the  by (\ref{eq:conditional-independence}) events at $p_A$ and $p_B$ will be independent, so that: 
\bea
\P_E(AB |i_A i_B) &=& \P_E(A | i_A i_B) \, \P_E(B | i_A i_B) \nn \\
&=& \P_E(A | i_A) \, \P_E(B | i_B) \label{eq:PE-indep}
\eea
where we can take the second step since we assume that since $p_A$ and $p_B$ are spacelike, the choice of $i_A$ (which we have occurring at $p_A$ itself without depending on any other spacetime region) does not affect the outcomes at $p_B$ nor does the choice of $i_B$ affect the outcomes at $p_A$. In other words $\P_E(A \, | \, i_Ai_B)  = \P_E(A \, | \, i_A)$ and $\P_E(B \, | \, i_Ai_B)  = \P_E(B \, | \, i_B)$.

Now in a flat spacetime we would expect the `classical' result of such a polarisation measurement \cite{aharonov:paradoxes}
\bea
\P_E(A^+ \, | \, \th_A) &=& cos^2\th_A \nn \\
\P_E(A^- \, | \, \th_A) &=& sin^2\th_A \label{eq:classical-measurement}
\eea
With a similar result at $p_B$. We will generalise this to:
\bea
\P_E(A | \th_A i_A) &=& f(A, \th_A) \nn \\
\P_E(B | \th_B i_B) &=& f(B, \th_B)
\eea
for some function $f$. We can now apply our decomposition of the holonomy (see section \ref{subsec:th-decomp}),
\beq
\th_A = \thav + \thao + \thaa
\eeq
Now although $\thao$ is known at $p_E$, both $\thaa$ and $\th_{\a}$ are unknown. However we are able to further decompose the event $\E_E(\th_{\a})$. The angle $\th_{\a}$ is determined by $v_{\a}$, $m_{\a}$ and $a^{i_A}_{/a}$, and while the former pair of variables are known at $p_E$ the later pair are not. We will need to go even further; since the potential measurement directions $\{a^i_{\a}\}$ are known at $p_E$, the only unknown variable is the choice $i_A$ of one of these directions, which we think of as being made at $p_A$. Then $i_A$ and $\thaa$ will fix $\th_A$. Conversely, it is easy to see that the combination of $\th_A$ and $i_A$ will fix $\thaa$. Then the event in $\L_E$ that both $\thaa$ and $i_A$ occur is equivalent to the event that both $i_A$ and $\th_A$ occur, so that:
\beq
\P_E(A \, | \, \th_Ai_A) = P_E(A \, | \, \thaa i_A)
\eeq
Then because we have assumed that the choice of $i_A$ is independent of all other events, we have:
\[
\P_E(A | i_A \thaa) = \frac{\P_E(A \thaa | i_A)}{\P_E(\thaa)}
\]
then,
\[
\P_E(A \thaa | i_A) = \P_E(\thaa) f(A,\th_A)
\]
so that,
\[
\P_E(A | i_A) = \int_0^{2\pi} \P_E(\thaa) f(A,\th_A) d\thaa
\]
with a similar result holding for $B$. Putting these together using (\ref{eq:PE-indep}) we get:
\beq\label{eq:PE-dyn}
\boxed{\P_E(AB |i_A i_B) = \Bigg(\int_0^{2\pi} \P_E(\thaa) f(A,\th_A) d\thaa\Bigg)\Bigg( \int_0^{2\pi} \P_E(\thbb) f(B,\th_B) d\thbb\Bigg)}
\eeq

\subsubsection{The General Dynamics}\label{subsec:chsh-po}

The next step is to shift our point of view from $\P_E$ back to $\P_O$. To do this, we make the simplifying assumption that the only variables affecting the measurement outcomes are the angles and vectors we have discussed, so that,
\bea
\P_E(AB |i_A i_B) &=& \P_O(AB \, | \, i_A i_B E) \nn\\
 &=&  \P_O(AB \, | \, i_A i_B \thv \thao \thbo) \nn \\
 &=& \frac{\P_O(AB \thv \thao \thbo\, | \, i_A i_B )}{\P_O(\thv \thao \thbo)}
\eea
We have assumed that the choice of $\thv$ is independent of all other variables other than the outcomes $A,B$, which means that
\beq
\P_O(\thv \thao \thbo) = \P_O(\thv)\P_O(\thao \thbo)
\eeq
We do not expect $\thao$ and $\thbo$ to be independent in general, so we can not separate this expression any further. Putting this together we get:
\beq
\P_O(AB \, | \, i_A i_B) = \int_0^{2\pi}\int_0^{2\pi}\int_0^{2\pi} \P_O(\thv)\P_O(\thao \thbo)\P_E(AB \, | \, i_A i_B) d\thv d\thao d\thbo
\eeq
Then using (\ref{eq:PE-dyn}) we achieve our general result:
\bea
\P_O(A B \, | \, i_A i_B) = 
\int_{\thao=0}^{2\pi}\int_{\thbo=0}^{2\pi}\P_O(\thao \thbo)\int_{\thv=0}^{2\pi} && \P_O(\thv)\ast \nn \\
&& \Bigg(\int_0^{2\pi} \P(\thaa) f(A, \th_A) \, d\thaa \Bigg)\ast \nn \\
&& \Bigg(\int_0^{2\pi} \P(\thbb) f(B, \th_B) \, d\thbb\Bigg) \nn \\
&& d\thv d\thao d\thbo  \label{eq:PO-dyn} 
\eea
 
\noindent Note that this will not in general separate into $A$ and $B$ probabilities as the $\P_E$ result does. This separation is crucial for the CHSH result, which is immediately thrown into question in the presence of curvature. However rather than compute the CHSH statistic we will instead make several simplifications leading to an even more interesting result.
 
\subsection{Simplification}\label{subsec:simplification}
 
 In this section we introduce several simplifying assumptions and apply them to our general result. All of our subsequent calculations will be based at $p_O$, so in what follows we will omit the $O$ index in the interests of increased readability. 
 
\subsubsection{Simplifying Assumptions}
 
We can now simplify (\ref{eq:PO-dyn}) by introducing some assumptions.

\begin{enumerate}
\item \label{as:Oflat} We assume that the spacetime is flat outside $\le$ so that $\thaa=\thbb=0$. 
\item \label{as:Ounif} We assume that $\thv$ is uniformly distributed, $\P(\thv)=1/2\pi$
\item \label{as:Odf} We assume that there is only a single degree of freedom in $\thao,\thbo$, which we can express as $\thm=\thao-\thbo$. Our motivation is the intuition that `non-classical' behaviour is driven by the difference between the effects of holonomy on the $A$ and $B$ measurements.
\item \label{as:f} We assume that $f$ takes the standard classical form given by (\ref{eq:classical-measurement}).
\end{enumerate}

\subsubsection{Simplification}

Now we can define:
\bea
F_{AB} &=& f(A,\th_A)f(B,\th_B) \nn \\
&=& f(A,\thav + \thao)f(B,\thbv - \pi + \thbo)
\eea
since we have assumed that $\thaa=\thbb=0$, and because $\th_O(b_O,-v_O) = \thbo-\pi$. Then applying assumption \ref{as:Oflat} our dynamics (\ref{eq:PO-dyn}) becomes:
\beq
\P(AB \, | \, i_A i_B) = \int_{\thb=0}^{2\pi}\int_{\tha=0}^{2\pi} \P(\thao\thbo)I_v(A,B)d\tha d\thb
\eeq
where:
\beq
I_v(A,B) = \int_{\thv=0}^{2\pi} \P(\thv) F_{AB} d\thv
\eeq
Applying assumption \ref{as:Ounif} we have $\P(\thv) = \sfrac{1}{2\pi}$ so that,
\beq
I_v(A,B) = \frac{1}{2\pi}\int_{\thv=0}^{2\pi} F_{AB} d\thv
\eeq
Then applying assumption \ref{as:Odf}, (\ref{eq:PO-dyn}) becomes:
\beq\label{eq:simp-dyn}
\boxed{\P(AB \, | \, i_Ai_B) =\int_{\thm=0}^{2\pi} \P(\thm)I_v(A,B)d\thm}
\eeq
Note that there is only one remaining unknown function, $\P(\thm)$. Finally, appendix \ref{appendix:int-calcs} steps through the calculations to show that applying assumption \ref{as:f} gives us:
\bea
I_v(+1,+1) = I_v(-1,-1) = \sfrac{1}{4} \, (\sfrac{1}{2} + cos^2\th_-) \label{eq:Ivpp}\\
I_v(+1,-1) = I_v(-1,+1) = \sfrac{1}{4} \, (\sfrac{1}{2} + sin^2\th_-) \label{eq:Ivpm}
\eea
where,
\bea
\th_- &=& \tha - \thb \nn \\
&=& \thab + \psi_- + \pi \label{eq:thm}
\eea

\subsubsection{Quantum Probabilities}

Our overall aim in this section is to assess the gap between Quantum (CH) and Classical (GR) behaviour as revealed by the CHSH framework. To complete this task, we examine whether we can find a solution of (\ref{eq:simp-dyn}) for $\P(\thm)$ which will yield the quantum probabilities $Q(AB \, | \, i_Ai_B)$ for this experiment. We set,
\beq\label{eq:qm}
\P(AB \, | \, i_Ai_B) = Q(AB \, | \, i_Ai_B)
\eeq
where,
\bea
&& Q(A^+B^+ \, | \, i_Ai_B) = Q(A^-B^- \, | \, i_Ai_B) = \sfrac{1}{2} \, sin^2 (\thab/2) \label{eq:qpp} \\
&& Q(A^+B^- \, | \, i_Ai_B) =  Q(A^-B^+ \, | \, i_Ai_B) = \sfrac{1}{2} \, cos^2 (\thab/2) \label{eq:qpm}
\eea
Notice that the $Q$'s follow the same pattern as our $I_v$ integrals. Then paring (\ref{eq:Ivpp}) with (\ref{eq:qpp}) and (\ref{eq:Ivpm}) with (\ref{eq:qpm}), substituting into (\ref{eq:simp-dyn}) yields:
\bea
\sfrac{1}{2} \, sin^2 (\thab/2) &=& \frac{1}{4} \int_0^{2\pi} \P(\thm) (\sfrac{1}{2} + cos^2\th_-) \, d\thm \label{eq:qms} \\
\sfrac{1}{2} \, cos^2 (\thab/2) &=& \frac{1}{4} \int_0^{2\pi} \P(\thm) (\sfrac{1}{2} + sin^2\th_-) \, d\thm \label{eq:qmc}
\eea
Now we can check that our probabilities are well defined, so that,
\beq
\sum_{A,B\in\{\pm 1\}} \P(AB \, | \, i_Ai_B) = \sum_{A,B\in\{\pm 1\}} Q(AB \, | \, i_Ai_B) = 1
\eeq
This acts as a constraint which makes (either) one of (\ref{eq:qms}), (\ref{eq:qmc}) redundant. Without loss of generality we will use (\ref{eq:qms}), and noting that $\int_0^{2\pi}\P(\thm)\,d\thm=1$ we can simplify by integrating over the constant term to yield,
\beq
2 \, sin^2 (\thab/2) - \sfrac{1}{2} = \int_0^{2\pi} \P(\thm) \, cos^2\th_- \, d\thm
\eeq
Finally we can simplify the LHS using $2 \, sin^2 (\thab/2) = (1-cos\thab)$ and (\ref{eq:thm}) to yield:
\beq\label{eq:qm}
\boxed{\sfrac{1}{2} - cos\,\thab = \int_0^{2\pi} \P(\thm) \, cos^2(\thab + \thm + \pi)\,d\thm}
\eeq
Thus a solution to (\ref{eq:qm}) for $\P(\thm)$ would allow us to derive the quantum dynamics from our `classical' framework.

\subsection{Discussion}\label{subsec:chsh-discussion}

Our aim in this section was to use the CHSH framework to identify and explore the gap between Quantum and Classical behaviour in our theory of GR observers. Unfortunately the CHSH framework fails to achieve this on a curved background, and if \ref{eq:qm} can be solved we may even be able to reconstruct the quantum dynamics in our classical system.

This occurs because of the joint dependency of the two measurements on fields (principally curvature) in the joint past of the measurement points. Indeed, upon conditioning on this region in section \ref{subsec:chsh-pe} we found the probability to factorise, which would lead directly to the CHSH result. 

Now the possibility of a correlation through dependency on the joint past is an obvious concern which was not overlooked by Bell and CHSH. For example it is discussed at length in the final chapter of \cite{bell:speakable} where Bell defines the concept of \emph{Local Causality} (LC), which is necessary for the CHSH inequality to hold. Indeed, Bell holds that the separability of the measure (which we found in the case of $\P_E$) is a consequence of local causality, and uses the adherence to the CHSH inequality as a test of compatibility with LC. It is argues that QM is not locally causal, and it is assumed without comment that a `classical theory' is. The Aspect experiment is then interpreted as finding against LC and `classical behaviour'. However, as we have seen, it turns out that the global nature of holonomy, and its particular relevance to the process of measurement, means that the actual \emph{measurements} we make in GR may not be formulated in a locally causal manner.

\section{Toward a Global Logical Structure}\label{sec:global}

In this section we take the initial steps toward constructing a global logical framework which would bring together the various worldviews of our spacetime observers into a single object. Due to the parallels of our framework with CH, we will sketch out how the topos techniques pioneered by Isham \cite{isham:toposcs} to achieve this goal in CH might be adapted to observer worldviews in GR.

\subsection{From Observers to Spacetime}

We begin by noting that our assumption in (\ref{eq:as-knowledge}) of total knowledge of the causal past implies directly that $\Ko_p$ depends only on the spacetime point $p$, and not the observer $\O$, so that:
\beq\label{eq:knowledge-indep}
K^{\O_1}_p = K^{\O_2}_p
\eeq
for all observers $K^{\O_1}$, $K^{\O_2}$ whose worldlines include $p$. Note that if we weaken our assumption of complete knowledge of the causal past (\ref{eq:as-knowledge}) we might still have this result, for example we might imagine that any two observers meeting at a point $p$ would `share' their knowledge of the past, leading directly to the above.

Further, it is clear from our construction of $\W_p$ and $\L_p$ in section \ref{subsec:prob-th} that given (\ref{eq:knowledge-indep}), both of these structures depend only on the spacetime point $p$ and not on the observer. 
\bea\
\W^{\O_1}_p &=& \W^{\O_2}_p \nn \\
\L^{\O_1}_p &=& \L^{\O_2}_p \label{eq:logic-indep}
\eea

Finally, the probability measure used at $p$ is not fixed by $\L_p$, and may vary from observer to observer. We can then require $\P^{\O_1}_p = \P^{\O_2}_p$, motivating this assumption by thinking of the observers as all using the same `theory' (though we will not here go into precisely what we might mean by this term), by which they draw the same conclusions from the same knowledge.

\subsubsection{The Logical Framework}\label{subsec:obs-logic}

In this section we consider how to put together the individual `logical frameworks' used by the observers at each spacetime point into a single global framework. To achieve this we employ the `varying set' construction introduced by Isham \cite{isham:toposcs} to put together the various `worldviews' of consistent histories quantum mechanics into a single structure. The use of this framework is a natural consequence of the similarity between observer dependence in general relativity and the consistent histories approach to quantum mechanics which we noted above.

As we have seen above, each observer $\O$ constructs a Boolean event algebra $\L^{\O}_p$ for every point $p$ on its worldline $\go$. As we have shown above (\ref{eq:logic-indep}), our assumptions imply that this algebra is depends only on the spacetime point (and not the observer), so we have a unique Boolean algebra $\L_p$ for every spacetime point which is on the worldline of some observer. Noting that we have not restricted the set of worldlines which may be considered as observers, we will make the simplifying assumption that every spacetime point is on at least one observer's worldline, so that we have a unique algebra $\L_p$ $\forall p\in\M$. Then writing $\L(\M)=\{\L_p \, | \, p\in\M\}$ and recalling section \ref{subsec:prob-theory-causality}, we see that both $\M$ and $\L(\M)$ are posets, the former ordered by causality and the later by inclusion as a subalgebra. We have seen above \ref{eq:event-algebra-causality} that these two orders are related,
\[
p\leq q \Rightarrow \L_p \geq L_q
\]

Then regarding the posets $\M$ and $\L(\M)$ as categories in the standard manner, the relationship $\L:p\longrightarrow \L_p$ defines a functor between $\M$ and $\L(\M)$. This construction is analogous to the space of Boolean subalgebras of a consistent histories orthoalgebra introduced by Isham \cite{isham:toposcs}, so intuitively we can think of observer dependence in `classical' spacetime as analogous to what Isham calls the `many world-views' picture of consistent histories quantum mechanics. To explore this point further, it may be interesting to `invert' Isham's argument, and attempt to construct a consistent histories style orthoalgebra (or Boolean manifold) from our functor; however this lies beyond the scope of our current discussion. Following rather than inverting Isham's argument, by defining sieve's as in definition 3.2 of \cite{isham:toposcs} we can write $\W(L_p)$ to denote the set of sieves at $L_p$, and can use the $\W(L_p)$ (which will be Heyting algebras \cite{isham:toposcs}) to furnish us with logical structures at each $p$. These Heyting algebras can then be related directly with our spacetime, 
\beq
\W(p) = \W(\L_p)
\eeq
and we can think of $\W$ as a functor from $\M$ to the category of sets (or as a `varying set over $\M$' in Isham's terminology). We can then build global propositional structures using the same techniques employed when these structures represent a consistent histories theory. A further examination of this area is beyond the scope of this discussion.

\section{Conclusion}

Our aim in this paper was to construct a theory of observer dependency in General Relativity, and to contrast this with Consistent Histories. We built an initial version of such a theory in section $2$, defining our observers in $2.1$ and formulating their worldviews in $2.2$, including the imposition of consistency conditions in $2.2.3$ which relate the relations between these worldviews to the causal structure of the background spacetime. In $2.3$ we found our structures to bear close resemblance to those used in CH, with a notion of `incompatibility' between worldviews present in GR as well as CH. This leads to the initial steps we took in section $4$ to apply topos techniques designed for CH to bring together the various worldviews into a single logical structure.

However the GR conception of incompatibility seemed weaker than its Quantum analogue, and in section $3$ we adapted the CHSH framework to a curved background so we could better identify the gap between the classical and quantum formulations of observer dependence. Unfortunately, having been designed for a flat background the CHSH framework turns out to be insufficient for this task, with the global nature of holonomy and its relevance to the local measurement processes meaning that actual measurement process in GR may not always be `locally causal'. We are able to violate the inequality with a classical theory on a curved background using a detailed account of the measurement process, and if (\ref{eq:qm}) can be solved might even be able to derive the quantum dynamics from GR.

\section{Acknowledgements}

The author thanks Wajid Mannan for his comments on the article, Katerina Nissan for her contribution to the diagrams, and the British Library for the use of their facilities.

\newpage
\appendix
\section{Detailed Decomposition of $\th_A$ and $\th_B$}\label{appendix:thA-thB-decomp}

In this appendix we work through the details of the decompositions of $\th_A$ and $\th_B$ in terms of holonomy to prove the results quoted in (\ref{eq:thA-thB}). We assume the notation of section \ref{subsec:th-decomp}.

We proceed by chasing our vectors around the various loops, $\g_{OEA}, \g_{OE\a}$  and $\g_{\a EA}$. Our observers will make their predictions from $p_O$, so we will take this as the `basepoint' for comparison. Since $p_O$ is not in $\g_{\a EA}$ we will use $p_{\a}$ as the reference point for this loop. We will translate $a_A$ `back' to $p_{\a}$ and $p_O$ along $\g_{OA}$, while we will send $v_E$ around the loops based at the two points.

We start by defining,
\bea\label{eq:vOaO}
v_O &=& \G_{EO}v_E \nn \\
a_O &=& \G_{AO}a_A
\eea

Now for any point $p$ along $\g_{OA}$ we define $a_p$ and $m_P$ as the translates of $a_O$ and $m_O$ along $\g_{OA}$. Then for a tangent vector $X_p$ at $p$ we define $\th_p(X_p,a_p)$ to be the angle between $a_p$ and the projection of $X_p$ onto the (two dimensional) $m_p$. In the case that $X_p$ is orthogonal\footnote{In the analysis of the dynamics below this outcome will generally have zero probability} to $m_p$ we will set $\th_p$ to be $\pi/2$. Note critically that $\th$ is preserved by parallel translation, which is a rigid transformation of the entire space.

Then following (\ref{eq:vOaO}) we define:
\beq
v_{\a} = \G_{E\a} v_E
\eeq 
Now we can measure the effects of holonomy around the three loops by defining:
\bea
\th_{OEA} &=& \th_O(\G_{AO}\circ\G_{EA}\circ\G_{OE} (v_{O}),a_O) - \th_O(v_O,a_O) \nn \\
\th_{OE\a} &=& \th_O(\G_{\a O}\circ\G_{E\a}\circ\G_{OE} (v_{O}),a_O) - \th_O(v_O,a_O) \nn \\
\th_{\a EA} &=& \th_{\a}(\G_{A\a}\circ\G_{EA}\circ\G_{\a E} (v_{\a}),a_{\a}) - \th_{\a}(v_{\a},a_{\a}) 
\eea
Note that:
\beq
\G_{AO}\circ\G_{EA}\circ\G_{OE} (v_O) = v_A
\eeq
Similarly,
\bea
\G_{\a O}\circ\G_{E\a}\circ\G_{OE} (v_{O}) &=& v_{\a}, \nn \\
\G_{A\a}\circ\G_{EA}\circ\G_{\a E} (v_{\a}) &=& v_A
\eea
Then since parallel translation preserves $\th$ we have:
\bea
\th_O(\G_{AO}\circ\G_{EA}\circ\G_{OE} (v_{O}),a_O) &=& \th_A( v_{A},a_A) \nn \\
&=& \th_A
\eea
Similarly, writing $\th_{\a}=\th_{\a}(v_{\a},a_{\a})$ and $\thav=\th_{O}(v_{O},a_{O})$ we have:
\bea
\th_O(\G_{\a O}\circ\G_{E\a}\circ\G_{OE} (v_{O}),a_O) =\th_{\a}, \nn \\
 \th_{\a}(\G_{A\a}\circ\G_{EA}\circ\G_{\a E} (v_{\a}),a_{\a}) = \th_A
\eea
So that,
\bea\label{eq:th-defs}
\th_{OEA} = \th_A - \thav \nn \\
\th_{OE\a} = \th_{\a} - \thav \nn \\
\th_{\a EA} = \th_A - \th_{\a}
\eea
Therefore we can decompose the effects of holonomy,
\beq
\th_{OEA} = \th_{OE\a} + \th_{\a EA}
\eeq
which gives us,
\beq
\th_A = \thav + \th_{OE\a} + \th_{\a EA}
\eeq
Finally, to avoid cluttering our calculations we will slightly abbreviate our notation. Thinking of the loop $\g_{OE\a}$ as `loop $A_1$' and $\g_{\a EA}$ as `loop $A_2$' we will write:
\bea
\thao &=& \th_{OE\a} \nn \\
\thaa &=& \th_{\a EA}
\eea
so that:
\beq\label{eq:thA}
\th_A = \thav + \thao + \thaa
\eeq
We can perform a similar decomposition for $\th_B$, appointing a point $p_{\b}$ in the same manner as $p_{\a}$, so that:
\beq\label{eq:thB}
\th_B = \thbv + \thbo + \thbb
\eeq
where $\thbv = \th_O(b_O,v_O) = \th_O(b_O,-v_O) - \pi$, and we define $\thbo, \thbb$ in the same manner as $\thao, \thaa$.

\newpage
\section{Integral Calculations}\label{appendix:int-calcs}

In this appendix we expand upon the calculations behind (\ref{eq:Ivpp}) and (\ref{eq:Ivpm}) in section \ref{subsec:simplification}, whose notation we will assume.

Now in the notation of section \ref{subsec:simplification} we wish to determine
\beq
I_v(A,B) = \frac{1}{2\pi}\int_{\thv=0}^{2\pi} F_{AB} \, d\thv
\eeq
where
\bea
F_{AB} &=& f(A,\th_A)f(B,\th_B) \nn \\
&=& f(A,\thav + \thao)f(B,\thbv - \pi + \thbo)
\eea
and
\bea
P_E(A^+ \, | \, \th_A) &=& cos^2\th_A \nn \\
P_E(A^- \, | \, \th_A) &=& sin^2\th_A 
\eea
Putting these together we have:
\bea
F_{++} &=& cos^2 \th_A \, cos^2 \th_B \nn \\
F_{+-} &=& cos^2 \th_A \, sin^2 \th_B \nn \\
F_{-+} &=& sin^2 \th_A \, cos^2 \th_B \nn \\
F_{--} &=& sin^2 \th_A \, sin^2 \th_B
\eea
Then writing:
\bea
\th_+ &=& \th_A + \th_B \nn \\
\th_- &=& \th_A - \th_B
\eea
we have:
\bea
F_{++} &=& \sfrac{1}{4}(cos^2\th_+ + 2 cos \th_+ \, cos \th_- + cos^2\th_-) \nn \\
F_{--} &=& \sfrac{1}{4}(cos^2\th_+ - 2 cos \th_+ \, cos \th_- + cos^2\th_-) \nn \\
F_{+-} &=& \sfrac{1}{4}(sin^2\th_+ - 2 sin \th_+ \, sin \th_- + sin^2\th_-) \nn \\
F_{-+} &=& \sfrac{1}{4}(sin^2\th_+ - 2 sin \th_+ \, sin \th_- + sin^2\th_-) \label{eq:F}
\eea
To integrate these expressions we begin by making their dependence on $\thv$ more explicit. This is achieved by an examination of the angles $\th_+$, $\th_-$:
\bea
\th_+ &=& \th_A + \th_B \nn \\
&=& \thav + \thbv + \thao + \thbo - \pi \nn \\
&=& -2\thv +  \overbrace{\tha + \thb + \thao + \thbo - \pi}^{=\phi_+}
\eea
and,
\bea
\th_- &=& \th_A - \th_B \nn \\
&=& \thav - \thbv + \thao - \thbo + \pi \nn \\
&=& \thab + \psi_- + \pi
\eea 
Since $\thav - \thbv = \th_O(a_O,v_O) - \th_O(b_O,v_O) = \th_O(A_O,B_O) = \thab$. Note critically that \emph{$\th_-$ has no dependence on $\thv$}.

We are now in a position to integrate the terms within (\ref{eq:F}). For example, looking at the $cos^2$ terms we have:
\bea
\frac{1}{2\pi}\int_{\thv=0}^{2\pi}cos^2\th_+ \, d\thv &=& \frac{1}{4\pi}\int_{\thv=0}^{2\pi}(1+cos2\th_+) \, d\thv \nn \\
&=& \frac{1}{4\pi}\underbrace{\big[\thv\big]_0^{2\pi}}_{=2\pi} + \cancelto{0}{\frac{1}{4\pi}\int_{\thv=0}^{2\pi}cos(-4\thv + \phi_+) \, d\thv} \nn \\
&=& \frac{1}{2}
\eea
and,
\bea
\frac{1}{2\pi}\int_{\thv=0}^{2\pi}cos^2\th_- \, d\thv &=& \frac{cos^2\th_-}{2\pi}\overbrace{\big[\thv\big]_0^{2\pi}}^{2\pi} \nn \\
&=& cos^2\th_-
\eea
Similarly we get:
\bea
\frac{1}{2\pi}\int_{\thv=0}^{2\pi}cos\th_+ \, cos\th_- \, d\thv &=& 0 \nn \\
\frac{1}{2\pi}\int_{\thv=0}^{2\pi}sin^2\th_+ \, d\thv &=& \frac{1}{2} \nn \\
\frac{1}{2\pi}\int_{\thv=0}^{2\pi}sin\th_+sin\th_- \, d\thv &=& 0 \nn \\
\frac{1}{2\pi}\int_{\thv=0}^{2\pi}sin\th_+ \, d\thv &=& sin^2\th_-
\eea
Putting this all together we have:
\bea
I_v(+1,+1) = I_v(-1,-1) = \sfrac{1}{4} \, (\sfrac{1}{2} + cos^2\th_-) \\
I_v(+1,-1) = I_v(-1,+1) = \sfrac{1}{4} \, (\sfrac{1}{2} + sin^2\th_-) 
\eea

\newpage
\bibliography{Bib}
\bibliographystyle{plain}

\end{document}